%% file: main_submission.tex
\documentclass{article}
\usepackage{graphicx}
\usepackage{amsmath,amssymb}
\usepackage{geometry}
\usepackage{booktabs}
\usepackage{authblk}

\usepackage{pgfplots}

\usepackage{siunitx}
\usepackage[unicode]{hyperref}
\usepackage{subcaption}
\pgfplotsset{compat=newest}

\usepackage{tikz}
\usepackage{tikz-feynman}
\tikzfeynmanset{
  compat=1.1.0,
  hadblob/.style={
    circle,
    draw,
    fill=cyan!20,
    minimum size=1.5em,
    inner sep=0pt
  }
}
\usetikzlibrary{decorations.pathmorphing,positioning}
\usepackage[utf8]{inputenc}
\usepackage{textgreek}

\begin{document}

%\title{Updated $(g-2)_\mu$ and $(g-2)_e$ allowed parameter space for heavy photons and comparisons to the ATOMKI X17 Measurement}
\title{Compatibility of the Updated $(g-2)_\mu$, $(g-2)_e$ and PADME-Favored Couplings with the Preferred Region of ATOMKI X17}%, Given a Neutrino Suppressed and SU2 violating Protophobic Vector Interpretation}

%------------------------------------------------------------
% Author and affiliation block (requires \usepackage{authblk})
%------------------------------------------------------------
% Put authors on one line, separated by commas
\renewcommand\Authsep{, }      % between authors
\renewcommand\Authands{, }     % before the last author (no “and”)
\renewcommand\Authand{, }      % two-author case
\setlength{\affilsep}{2pt}     % tighter gap above affiliations (optional)

% ----------  AUTHORS  ----------
\author[1,2]{Emrys Peets\thanks{Corresponding author: \texttt{epeets@stanford.edu}}}

\affil[1]{Department of Physics, Stanford University, Stanford, CA 94305, USA}
\affil[2]{Fundamental Physics Directorate, SLAC National Accelerator Laboratory, Menlo Park, CA 94025, USA}

%\date{\today}
% Footnote for corresponding author
\renewcommand{\thefootnote}{\fnsymbol{footnote}}

\maketitle

\begin{abstract}
We re-evaluate the viability of a kinetically mixed dark photon ($A^{\prime}$) as a solution to the muon anomalous magnetic moment $(g-2)\mu$ discrepancy and the ATOMKI nuclear anomalies near 17~MeV, using the final FNAL measurement and the latest theory predictions (BMW21, WP25). For $m_{A^{\prime}} = 17$~MeV, the allowed kinetic mixing parameter narrows to $\varepsilon_\mu = 7.03(58)\times10^{-4}$ (WP25). We then directly compare the allowed region for the muon and X17 bands to those preferred by the electron magnetic moment measurements. For the electron, we obtain $\varepsilon_e = 1.19(15)\times10^{-3}$ (Cs, 2018) and $\varepsilon_e = 0.69(15)\times10^{-3}$ (Rb, 2020), based on two recent measurements of the fine structure constant compared to the most recent experimental value determined using a one-electron quantum cyclotron. This study focuses on the protophobic vector interpretation of X17 and assumes $\varepsilon_\nu<<\varepsilon_l$. While a mild tension persists, we identify a narrow overlapping region, $6.8\times10^{-4} \lesssim \varepsilon \lesssim 9.6\times10^{-4}$, between recent PADME results, the NA64 exclusion, and within the 2$\sigma$ preferred coupling region given by the Rb 2020 determination of $\alpha_\varepsilon$. These results provide well-defined targets for future experimental searches and motivate further theoretical refinements, both of which will play a decisive role in assessing the validity of the ATOMKI anomaly claims. Of particular note is the fixed target X17 experiment to be conducted in Hall-B of Thomas Jefferson National Accelerator Facility in Summer of 2026.

\end{abstract}

\section{Introduction}
% --- Copy/paste-ready intro text (2 paragraphs) + key equations ---

Light, weakly-coupled spin-1 mediators arise naturally in extensions of the Standard Model by an
additional $U(1)$ gauge factor, where the associated gauge boson $X_\mu$ can interpolate between a
dark-photon-like state (coupled dominantly to $Q$ via kinetic mixing), a dark-$Z$-like state (via $Z$--$X$
mixing), and more general forces coupled to linear combinations of $Q$, $B$, and $L$ (or $B\!-\!L$).
In a generic low-energy effective description one may write
\begin{equation}
\mathcal{L}\ \supset\ -\frac14 X_{\mu\nu}X^{\mu\nu}+\frac12 m_X^2 X_\mu X^\mu
\;+\;X_\mu\sum_f \bar f\gamma^\mu\!\left(g_V^f+g_A^f\gamma^5\right)f\,,
\label{eq:genericXlag}
\end{equation}
where $g_V^f$ and $g_A^f$ are the vector and axial couplings to SM fermions.
In ultraviolet completions with extra Higgs structure (e.g. multiple doublets plus singlets), the
effective $U(1)$ current can contain both vector and axial pieces, and the vector part may be arranged
to be approximately ``protophobic''---i.e. the effective proton coupling is suppressed compared to the
neutron coupling \cite{Fayet1986FifthInteraction,Fayet1989FifthForceCharge,Fayet1990ExtraU1,Fayet2017LightU}.
At the quark level this is expressed as
\begin{equation}
g_V^p \equiv 2g_V^u+g_V^d,\qquad g_V^n \equiv g_V^u+2g_V^d,\qquad |g_V^p|\ll |g_V^n|\,,
\label{eq:protophobic}
\end{equation}
with a particularly natural realization from $Z$--$X$ mixing yielding a small ratio $g_V^p/g_V^n$
(set by electroweak charges) \cite{Fayet1986FifthInteraction}.
Such a protophobic, MeV-scale vector has been widely discussed in connection with the ATOMKI
$e^+e^-$ angular-correlation anomalies \cite{Krasznahorkay2016Be8,Feng2016Protophobic}.
In this work we focus on the protophobic-vector interpretation and adopt the neutrino-suppressed
benchmark $|g_{\nu}|\ll |g_e|$ (equivalently $|\epsilon_\nu|\ll|\epsilon_e|$ in mixing language), as motivated by
neutrino-scattering constraints  \cite{DentonGehrlein2023Neutrino}.

A central phenomenological handle on such light vectors is the one-loop contribution to lepton
magnetic moments, which cleanly separates vector and axial pieces and highlights why axial couplings
are typically much more tightly constrained at low mediator mass \cite{Fayet2007UbosonProduction}.
Writing the charged-lepton couplings as $f_{\ell V}\equiv g_V^\ell$ and $f_{\ell A}\equiv g_A^\ell$, one finds model contributions to $a_\ell$ of
\begin{equation}
\delta a_\ell\ \simeq\ \frac{f_{\ell V}^2}{8\pi^2}\,G\!\left(\frac{m_X}{m_\ell}\right)\;-\;
\frac{f_{\ell A}^2}{4\pi^2}\,\frac{m_\ell^2}{m_X^2}\,H\!\left(\frac{m_X}{m_\ell}\right),
\label{eq:g2VA}
\end{equation}
with loop functions that may be written as the Feynman-parameter integrals
\begin{align}
G(r) &= 2\int_0^1\!dx\;\frac{x^2(1-x)}{x^2+r^2(1-x)}\,, \\
H(r) &= \int_0^1\!dx\;\frac{2x^3+(x-x^2)(4-x)\,r^2}{x^2+(1-x)\,r^2}\,,
\end{align}
so that for $r\ll1$ one has $G(r)\to 1$ and $H(r)\to 1$ while the axial term is enhanced by the
explicit factor $m_\ell^2/m_X^2$ \cite{Fayet2007UbosonProduction}.
In the dark-photon limit (purely vector, flavor-universal coupling) one may identify
$f_{\ell V}=e\,\epsilon$ and $f_{\ell A}=0$, recovering the familiar $\delta a_\ell\propto \alpha\,\epsilon^2$
scaling used throughout dark-sector phenomenology \cite{Pospelov2009SecludedU1,Fayet2017LightU}.
This tight interplay between $(g-2)_\ell$, nuclear-transition kinematics, and protophobic charge assignments
motivates targeted tests of the remaining MeV-scale parameter space.

\subsection*{This Study}
Using the latest results from the (g-2) experiment \cite{MuonGminus2:2025Run1to6}, and considering BMW lattice QCD corrections to the gyromagnetic ratio to the muon \cite{Borsanyi:2020mff}, we report updated allowed parameter space for dark sector heavy photons between masses of 5 and 500 MeV that could couple to muons. We include the first comparison of the $(g-2)_\mu$ allowed $\varepsilon$ and the preferred coupling given the ATOMKI measurements \cite{Krasznahorkay:2015iga, Krasznahorkay:2019lyl,PhysRevLett.117.071803}. We illustrate how the theoretical prediction of the magnetic moment has changed over time by comparing to the 2020 G-2 white paper, and the 2021 BMW correction including lattice QCD corrections \cite{Aoyama:2020ynm, Borsanyi:2020mff}. 

Additionally, we include a comparison with the allowable coupling of a heavy photon calculated from $(g-2)_e$ using precision measurements of $\alpha$ from Cs and Rb measurements, noting small exclusions of the preferred ATOMKI coupling solely from the two measurments \cite{Parker:2018vye, Morel:2020dww}. This study discusses the viability of a vector boson at the ATOMKI mass and notes that only a sliver of overlap exists between current experimental constraints and a protophobic vector boson within the preferred ATOMKI region\cite{PhysRevLett.117.071803}. The model of interest for a vector boson must also have neutrino suppression. This study focuses on this vector model interpretation, though the author recognizes the pseudo-scalar interpretation has not been fully excluded in the ATOMKI favored region. For the purposes of this study, we do not overlay all experimental exclusions and visually only compare the favored ATOMKI bounds to allowed $\varepsilon_\ell$ from each $(g-2)_\ell$, the reported PADME upperlimit and the NA64 exclusion contour.

\section{Current Values of the $\Delta a_\mu$ and $\Delta a_e$}

\subsection{Latest Experimental Measurement and Standard Model Predictions for $a_\mu$}

The anomalous magnetic moment of the muon, 
\begin{equation}
  a_\mu \equiv \frac{g_\mu - 2}{2}.
\end{equation}

 is a stringent test of the Standard Model (SM) and a sensitive probe for new physics. Recent advances in both experiment and theory have led to an updated assessment of the longstanding discrepancy $\Delta a_\mu$ between measurement and the SM expectation. In this section, we summarize the latest values for $a_\mu$ and their uncertainties as of 2025, including both data-driven and lattice QCD evaluations of the hadronic vacuum polarization (HVP) contribution.

\subsection*{Discrepancy and Uncertainty Propagation}

The discrepancy between experiment and theory is calculated as
\begin{equation}
    \Delta a_\mu = a_\mu^{\rm exp} - a_\mu^{\rm SM}
\end{equation}
with the uncertainty calculated by adding in quadrature:
\begin{equation}
    \sigma_{\Delta a_\mu} = \sqrt{\sigma_{\rm exp}^2 + \sigma_{\rm SM}^2}
\end{equation}
\subsubsection*{FNAL Final Experimental Average} 
    %\textbf{OLD}
    %\begin{equation}
    %    a_\mu^{\rm exp} = 116\,592\,059\,(22) \times 10^{-11}
    %\end{equation}

    As reported by the Muon $g-2$ Collaboration , the final combined experimental (exp) average, from BNL E821 4 and Runs 1-6 \cite{MuonGminus2:2025Run1to6}.

    \begin{equation}
        a_\mu^{\rm exp} = 116\,592\,0715\,(145) \times 10^{-11}
    \end{equation}

    %\textbf{NEW}
    %1165920705 (205)
    
    \subsubsection*{Standard Model (SM) Predictions of $a_\mu$ and Corresponding $\Delta a_\mu$}
    \begin{itemize}
        \item \textbf{Dispersive/Data-driven HVP (Muon $g-2$ Theory Initiative, 2020)} To demonstrate how only the dispersive hadronic vacuum polarization influences the allowable parameter space that a dark, or heavy photon, we include the theory estimate from the 2020 (g-2) white paper \cite{Aoyama:2020ynm} 
        \begin{equation}
            a_\mu^{\rm SM,WP20} = 116\,591\,810\,(43) \times 10^{-11}
        \end{equation}

\begin{align}
    \Delta a_\mu^{\rm WP20} &= a_\mu^{\rm exp} - a_\mu^{\rm SM,WP20} \\
    &= [116\,592\,071.5 - 116\,591\,810] \times 10^{-11} \\
    &= 261.5 \times 10^{-11} \\
    \sigma_{\Delta a_\mu^{\rm WP20}} &= \sqrt{(14.5)^2 + 43^2} \times 10^{-11} \\
    &= \sqrt{210.25 + 1849} \times 10^{-11} \\
    &\approx 45.4 \times 10^{-11}
\end{align}

Thus,
\begin{equation}
    \boxed{
    \Delta a_\mu^{\rm WP20} = 262(45) \times 10^{-11}
    }
\end{equation}

        \item \textbf{BMW Lattice QCD HVP (BMW Collaboration, 2021)} Including the 2021 calculations from the BMW collaboration, highlight the significant shrinking of allowed parameter space
        \begin{equation}
            a_\mu^{\rm SM,BMW} = 116\,591\,954\,(67) \times 10^{-11}
        \end{equation}

\begin{align}
    \Delta a_\mu^{\rm BMW} &= a_\mu^{\rm exp} - a_\mu^{\rm SM,BMW} \\
    &= [116\,592\,071.5 - 116\,591\,954] \times 10^{-11} \\
    &= 117.5 \times 10^{-11} \\
    \sigma_{\Delta a_\mu^{\rm BMW}} &= \sqrt{(14.5)^2 + 67^2} \times 10^{-11} \\
    &= \sqrt{4699.25} \times 10^{-11} \\
    &\approx 68.6 \times 10^{-11}
\end{align}

Thus,
\begin{equation}
    \boxed{
    \Delta a_\mu^{\rm BMW} = 118(69) \times 10^{-11}
    }
\end{equation}

        \item 
        \textbf{Muon $g-2$ Theory Initiative, 2025:}
        Finally we use the latest theoretical predictions from the (G-2) theory initiative to create the latest allowable heavy photon bands.
        \begin{equation}
            a_\mu^{\rm SM,GWP} = 116\,592\,033\,(630) \times 10^{-11}
        \end{equation}

\begin{align}
    \Delta a_\mu^{\rm WP25} &= a_\mu^{\rm exp} - a_\mu^{\rm SM,GWP} \\
    &= [116\,592\,071.5 - 116\,592\,033] \times 10^{-11} \\
    &= 38.5 \times 10^{-11} \\
    \sigma_{\Delta a_\mu^{\rm WP25}} &= \sqrt{(14.5)^2 + 62^2} \times 10^{-11} \\
  &= \sqrt{4054.25}\,\times 10^{-11} \\
  &\approx 63.7 \times 10^{-11}
\end{align}

Thus,
\[
\boxed{
\Delta a_\mu^{\rm WP25} = 39(64) \times 10^{-11}
}
\]

        %\item It is possible to include a combination of $a_\mu^{WP25}$ with the 2024 BMW calculations based on improved lattice models, but this will be done in later versions of this document. 
    \end{itemize}

\subsection{Anomalous Magnetic Moment of the Electron}

\input{electron_values}

%\noindent \textbf{Summary:}
%\begin{align}
%    \Delta a_\mu^{\rm disp} &= 249(48) \times 10^{-11} \\
%    \Delta a_\mu^{\rm BMW} &= 105(71) \times 10^{-11}
%\end{align}
%with uncertainties added in quadrature.

\input{heavy_photon_mixing}

\section{Current Feasibility of a 17 MeV Vector Boson}

\subsection{ATOMKI Measurements}
The ATOMKI Collaboration first reported a $> 5\sigma$ anomaly in the internal pair-creation angular correlations for the 17.6 MeV M1 transition in $^8$Be, which can be interpreted as the emission of a new vector boson X17 with mass $m_{A^\prime} = (16.7 \pm 0.85)$ MeV \cite{Krasznahorkay:2015iga}. This collaboration repeated these studies and then performed additional experiments focused around the $e^+e^-$ pair production of $^4$He and $^{12}$C nuclei. Respectively, the authors found similar excesses at large correlation angles with best-fit mass measurements of (17.01 $\pm $ 0.16) MeV, (16.94 $\pm$ 0.33 $MeV$), and (17.01 $\pm$ 0.31) MeV yielding a preferred kinetic mixing parameter in the range $\varepsilon_e \approx \times 10^{-4}-10^{-3}$. \cite{Krasznahorkay:2019lyl, PhysRevC.104.044003, PhysRevC.106.L061601}. For the purposes of this study, we consider the original ATOMKI mass of 16.7 MeV in the preferred range for $\varepsilon \in [2\times 10^{-4} , \;1.3 \times 10^{-3}]$ and preferred coupling to this as determined in \cite{PhysRevLett.117.071803}. 

\subsection{Additional Experimental Constraints at 17 MeV}

\subsection*{NA48/2 Exclusion and Protophobic Implication}

The NA48/2 limit from the decay chain
\[
\pi^0 \to \gamma A' \to \gamma e^+ e^-
\]
imposes
$\varepsilon^2 \lesssim 10^{-7} \quad\Rightarrow\quad \varepsilon \lesssim 3\times10^{-4}$ at \(m_{A'} \simeq 16.7~\text{MeV}\).
Therefore, any viable explanation of the ATOMKI excess must invoke a \emph{protophobic} gauge boson:
its effective coupling to protons (the relevant linear combination of $u$ and \(d\) quark charges) is suppressed to evade the \(\pi^0\) decay bound while retaining a sufficiently large coupling to electrons to generate the observed internal pair–creation anomaly\cite{PhysRevLett.117.071803}.

%
%Thus, though NA48/2 has the most stringent bounds on $\varepsilon_{\ell}$, a protophobic gauge boson could satisfy these bounds as discussed in \cite{PhysRevLett.117.071803} if within the sliver of parameter space between the PADME 95\% interval bounds and the NA64 90\% interval. This corresponds to a protophobic gauge boson allowable coupling range within the ATOMKI preferred region of $3.4\times10^{-4}\;\lesssim\;\varepsilon\;\lesssim\;5.6\times10^{-4}$. 

\subsection*{NA64 at CERN and Neutrino Suppression}

Complementary probes have already excluded large portions of this parameter space: the NA64 fixed‐target search at CERN rules out $1.3\times10^{-4} \leq \varepsilon_e \leq 6.8 \times 10^{-4}$ at 90\% CL for $m_{A^{\prime}}$ = 16.7 MeV \cite{NA64:2018}, while beam‐dump experiments (E141, E774), $e^+e^-$collider searches at BaBar and KLOE, and analyses by HADES, PHENIX, and NA48/2 constrain $\varepsilon^2\lesssim 10^{-7}$ for $9<m_{A'}<70,$MeV.

%--------------------------------------------------------
%\subsubsection*{Why the NA64 visible--decay limit shrinks for a
%\emph{protophobic} vector boson}

\subsubsection*{Neutrino couplings and the applicability of NA64 limits.}

\label{sec:ProtophobicRescale}
%--------------------------------------------------------

For the 2017--2018 combined datasets, NA64 reports a $90\%$ C.L.\ exclusion at
\begin{equation}
   1.2\times10^{-4}\;<\;\varepsilon_{e}^{\rm\,(NA64)}\;<\;6.8\times10^{-4},
   \qquad m_{X}=16.7~\text{MeV},
   \label{eq:NA64bean}
\end{equation}
where $\varepsilon_{e}$ is the coupling of the new boson to electrons \cite{NA64:2019vis}.
%Both edges of the ``bean'' are obtained under the benchmark assumption
%${\rm BR}(X\!\to e^{+}e^{-})\simeq 1$.

The NA64 ``visible'' exclusion in Eq.~\eqref{eq:NA64bean} is obtained under the benchmark assumption ${\rm BR}(X\!\to e^{+}e^{-})=1$ \cite{NA64:2019}. In a generic protophobic-$X$ interpretation, however, the coupling to neutrinos is model-dependent. In particular, reactor CE$\nu$NS data place strong bounds on $\varepsilon_{\nu_e}$ at $m_X \simeq 17~{\rm MeV}$, requiring $|\varepsilon_{\nu_e}| \ll |\varepsilon_e|$ in viable ATOMKI-motivated scenarios. This implies that the low-energy lepton couplings cannot satisfy the naive SU$(2)_L$ relation between $\varepsilon_{\nu_e}$ and $\varepsilon_e$ (e.g.\ $2\varepsilon_{\nu_e}=\varepsilon_e$ for purely vector couplings), i.e.\ an effective SU$(2)_L$-relation violation in the lepton sector is required \cite{DentonGehrlein2023Neutrino}.

Motivated by this, in the remainder of this work we adopt a neutrino-coupling-suppressed benchmark,
\begin{equation}
   |\varepsilon_{\nu_\alpha}| \ll |\varepsilon_e| \, ,
   \qquad \alpha=e,\mu,\tau,
   \label{eq:nuSupp}
\end{equation}
and assume no additional light invisible states below $m_X/2$. In this regime ${\rm BR}(X\to \nu\bar{\nu})\simeq 0$, so the NA64 visible-decay exclusion Eq.~\eqref{eq:NA64bean} can be applied without further rescaling.

\paragraph{Optional rescaling for a nonzero invisible width.}
For completeness, if $X$ has an invisible partial width (e.g.\ into neutrinos or other invisible states),
\[
   R \equiv \Gamma_{\rm inv}/\Gamma_{e^{+}e^{-}},
   \qquad {\rm BR}_{ee} = \frac{1}{1+R},
   \qquad \Gamma_{\rm tot}=\Gamma_{e^{+}e^{-}}(1+R),
\]
then the high-$\varepsilon_e$ (prompt-decay) edge of the visible NA64 ``bean'' shifts approximately as
\begin{equation}
   \varepsilon_{e}^{\text{(vis)}} \;\to\;
   \frac{\varepsilon_{e}^{\text{(vis)}}}{\sqrt{1+R}} \, ,
   \label{eq:highEdgeGeneral}
\end{equation} 
While the low-$\varepsilon_e$ (long-lived) edge is essentially unchanged up to small
acceptance effects.

If ${\rm BR}_{\rm inv}$ is sizeable, NA64's missing-energy search for invisibly decaying dark photons provides an additional constraint \cite{NA64:2023inv}; reinterpreting that result in a given protophobic-$X$ model requires specifying ${\rm BR}_{\rm inv}$ and the promptness of the invisible decay. We do not apply this constraint in the benchmark neutrino-suppressed scenario of Eq.~\eqref{eq:nuSupp}.

\subsection*{2025 PADME Observed Upper Limit and Coupling Conversion}
Most recently, the PADME experiment performed a resonant search via $e^+e^-$ annihilation on a fixed target, scanning $\sqrt{s}$ = 16.4–17.4 MeV; the data showed a local upward fluctuation at $\sqrt{s} \approx$ 16.90 MeV with a global significance of $\approx1.8\sigma$ but no definitive signal, and set 90\% CL upper limits on $g_{ve} \leq 5.6 \times 10^{-4}$ near the X17 mass \cite{PADME:2025}.

\paragraph{PADME limit and coupling conversion.}
PADME quotes limits in terms of a direct vector coupling $g_{ve}$ defined by
\begin{equation}
   \mathcal{L}\supset g_{ve}\,X_\mu\,\bar e\gamma^\mu e \, .
   \label{eq:PADME_Lint}
\end{equation}
\cite{PADME:2025}
To compare with $\varepsilon_e$, we use $g_{ve}=e\,\varepsilon_e$, with
$e\equiv\sqrt{4\pi\alpha_{\rm em}}$, i.e.
\begin{equation}
   \varepsilon_e = \frac{g_{ve}}{e}.
   \label{eq:gve_to_eps}
\end{equation}

PADME reports that the observed 90\% C.L.\ upper-limit curve is weakest near $M_X=16.90~{\rm MeV}$ with a corresponding global significance of $\approx1.8\sigma$ and reaches $g_{ve}^{90}\simeq 5.6\times 10^{-4}$ \cite{PADME:2025},
corresponding to
\begin{equation}
   \varepsilon_e^{\rm (PADME)}(16.90~{\rm MeV})
   \;\lesssim\; \frac{5.6\times 10^{-4}}{e}
   \;\simeq\; 1.9\times 10^{-3}.
   \label{eq:PADME_eps}
\end{equation}

\paragraph{Unexcluded interval (NA64 visible + PADME).}
Combining Eqs.~\eqref{eq:NA64bean} and \eqref{eq:PADME_eps}, the coupling interval
\begin{equation}
   6.8\times10^{-4}\;\lesssim\;\varepsilon_e\;\lesssim\;1.9\times10^{-3}
   \qquad (m_X\simeq 16.9~{\rm MeV})
   \label{eq:window_NA64_PADME}
\end{equation}
is not excluded by the NA64 displaced-decay search and the PADME resonant-production limit. The lower endpoint corresponds to the high-$\varepsilon_e$ edge of the NA64 ``visible'' bean, where the NA64 acceptance rapidly drops for promptly decaying $X\to e^+e^-$. This region is further excluded by the Rb2020 measurments as will be shown in the results section.

\subsection*{The PRad-II/X17 collaboration and the Hall~B X17 search concept}
Some studies exist that have focused on PIONEER and Mu3e as promising experimental tests to probe the X17 parameter space \cite{Di_Luzio_2025}. Another exciting prospect is the planned Hall~B ``X17 Search Experiment'' at Thomas Jefferson National Accelerator Facility (JLAB). This experiment is being pursued by the PRad-II/X17 collaboration, with co-spokespersons including A.~Gasparian, R.~Paremuzyan, D.~Dutta, N.~Liyanage, Ch.~Peng, H.~Gao, and T.~Hague \cite{GasparianHPS2024}. The collaboration’s near-term physics goal is twofold: (i) to validate (or set an upper limit on) electroproduction of the putative $\sim 17$~MeV state reported by ATOMKI, and (ii) to search more broadly for hidden-sector intermediate states in the $m_X \in [3,60]$~MeV range that decay to $e^+e^-$ and/or $\gamma\gamma$ \cite{GasparianHPS2024}.

Experimentally, the strategy is a forward-angle, fully-exclusive invariant-mass resonance search ``bump hunt'' in $e^- + \mathrm{Ta} \to e^- +A^\prime +\mathrm{Ta}\  \to e^- + e^+e^- + \mathrm{Ta}$, using a very thin heavy target (baseline $\sim 1~\mu$m Ta) and detecting all final-state particles to tightly constrain kinematics \cite{GasparianHPS2024,ParemuzyanHPS2025}. The detector apparatus is a magnetic-spectrometer-free configuration derived from the PRad setup on the Hall~B space frame upstream of CLAS12, combining the high-resolution PbWO$_4$ inner section of the HyCal electromagnetic calorimeter with two GEM tracking layers \cite{GasparianHPS2024,ParemuzyanHPS2025}. In this design, the calorimeter provides precise energy measurements and triggering, while the GEMs enable charged/neutral discrimination and rejection of non-target backgrounds (e.g.\ tracks originating from beamline material), enabling powerful exclusivity selections (energy consistency and coplanarity) \cite{GasparianHPS2024,ParemuzyanHPS2025}. The collaboration reports a target-constrained invariant-mass resolution of order $\sigma_m \simeq 0.48~\mathrm{MeV}$ for an $m_X \simeq 17~\mathrm{MeV}$ hypothesis, and emphasizes a complementary ``unconstrained'' reconstruction as a cross-check that any peak-like excess originates from the target; the setup is not intended as a displaced-vertex search \cite{GasparianHPS2024}.

\subsection*{Recent status}
By the June 2025 collaboration update, PRad-II and X17 had passed an Experiment Readiness Review (ERR) in May (with recommendations/comments to be addressed), and a preliminary installation/run schedule was presented with PRad-II commissioning and running preceding an X17 running window in May--July 2026 \cite{ParemuzyanHPS2025}. The same update reports progress on beamline and detector components (including HyCal channel testing and GEM construction/shipping milestones) consistent with an on-track path to the planned run period \cite{ParemuzyanHPS2025}.

%--------------------------------------------------------

\input{results}

%\input{electron_values}

%\input{atomki_comparison_v2}

\input{best_conclusion}

\section{Acknowledgements}
The author thanks Joseph Bailey and Aidan Hsu for discussions on existing exclusions and assistance in finding contours. The author thanks Anthony Morales for the discussions in helping clarify allowable leptonic coupling modes. The author is grateful for the many fruitful discussions with Andrew McEntaggert on Lattice QCD and explicit analytic forms. The author also thanks Rory O'Dwyer for double-checking calculations. AI was used as a tool for typesetting and formatting; all scientific content was created and validated by the author.

\bibliography{refs}

\bibliographystyle{unsrt}

\end{document}

%% file: electron_values.tex
%\subsection{Electron Anomalous Magnetic Moment}
\subsection*{Determining $\Delta a_e$ from Precision $\alpha$ Measurements}
\label{sec:delta_ae}

The electron anomalous magnetic moment is defined as
\begin{equation}
  a_e \equiv \frac{g_e - 2}{2}.
\end{equation}
Experimentally, $a_e$ is known with extremely high precision from trapped--electron measurements (Penning trap) \cite{Hanneke:2008, Hanneke:2011}.  The Standard Model (SM) prediction can be written as
\begin{equation}
  a_e^{\rm SM}(\alpha) = a_e^{\rm QED}(\alpha) + a_e^{\rm had} + a_e^{\rm EW},
\end{equation}
where $a_e^{\rm QED}$ is the perturbative QED series known through five loops, expressed as
\begin{equation}
  a_e^{\rm QED}(\alpha) = \sum_{n=1}^{5} C_{2n}\!\left(\frac{\alpha}{\pi}\right)^{n},
\end{equation}
with coefficients $C_{2n}$ given in Ref.~\cite{Aoyama:2012, Aoyama:2019}.  The small hadronic vacuum polarization, light–by–light term $a_e^{\rm had}$ and electroweak term $a_e^{\rm EW}$ are also included (see the consolidated SM evaluation in \cite{Aoyama:2019}).  Because $a_e^{\rm QED}$ dominates and depends sensitively on $\alpha$, different independent determinations of the fine structure constant yield different SM predictions $a_e^{\rm SM}(\alpha_i)$, and hence different residuals
\begin{equation}
  \Delta a_e^{(i)} \equiv a_e^{\rm exp} - a_e^{\rm SM}(\alpha_i).
\end{equation}

We consider two recent, high-precision determinations of $\alpha$:
\begin{enumerate}
  \item The 2018  $^{133}$Cs recoil measurement: $\alpha^{-1}_{\rm Cs18} = 137.035\,999\,046(27)$ \cite{Parker:2018vye}.
  \item The 2020 $^{87}$Rb recoil measurement: $\alpha^{-1}_{\rm Rb20} = 137.035\,999\,206(11)$ \cite{Morel:2020dww}.
\end{enumerate}

Using a common set of higher-order QED, hadronic, and electroweak contributions (as compiled in \cite{Aoyama:2019}) and the experimental value, as determined by a one-electron quantum cyclotron, 

%this was the previous 2008 penning trap measurment
%\begin{equation}    
%a_e^{\rm exp} = 0.001\,159\,652\,180\,73(28) 
%\end{equation}
%\cite{Hanneke:2008, Hanneke:2011}  

\begin{equation}    
a_e^{\rm exp} = 0.001\,159\,652\,180\,59(13) 
\end{equation}
\cite{PhysRevLett.130.071801}, the two SM predictions differ slightly due to the distinct $\alpha$ inputs.  Propagating uncertainties (treating the $\alpha$ error, the experimental $a_e$ error, and the SM theory truncation / hadronic / EW errors in quadrature), one obtains the residuals:

%from the 2008 penning trap value of electron
%\begin{align}
%  \Delta a_e^{(\mathrm{Rb\,2018})} &= a_e^{\rm exp} - a_e^{\rm SM}(\alpha_{\rm Rb18})
%   = -88 (36) \times 10^{-10}, \label{eq:deltaaeRb}\\[4pt]
%  \Delta a_e^{(\mathrm{Cs\,2020})} &= a_e^{\rm exp} - a_e^{\rm SM}(\alpha_{\rm Cs20})
%   = 46 (30)\times 10^{-10}. \label{eq:deltaaeCs}
%\end{align}

\begin{align}
  \Delta a_e^{(\mathrm{Cs\,2018})} &= a_e^{\rm exp} - a_e^{\rm SM}(\alpha_{\rm Cs18})
   = -102 (26) \times 10^{-10}, \label{eq:deltaaeRb}\\[4pt]
  \Delta a_e^{(\mathrm{Rb\,2020})} &= a_e^{\rm exp} - a_e^{\rm SM}(\alpha_{\rm Rb20})
   = 34 (16) \times 10^{-10}. \label{eq:deltaaeCs}
\end{align}

For each $\alpha_i$ we compute the uncertainties as
\begin{equation}
  \sigma^2(\Delta a_e^{(i)}) = \sigma^2(a_e^{\rm exp}) + 
  \left(\frac{\partial a_e^{\rm SM}}{\partial \alpha}\right)_{\alpha_i}^2 \sigma^2(\alpha_i)
  + \sigma^2_{\rm th,res},
\end{equation}
where $\sigma_{\rm th,res}$ encompasses the residual uncertainties (higher-order QED coefficient uncertainties, hadronic, and electroweak inputs).  The derivative is dominated by the leading QED term:
\begin{equation}
  \frac{\partial a_e^{\rm SM}}{\partial \alpha} \simeq 
  \frac{1}{\pi} C_{2} + \frac{2}{\pi^2} C_{4}\alpha + \cdots,
\end{equation}
with $C_2 = \tfrac{1}{2}$, $C_4 \approx 0.328478965\dots$ etc.\ \cite{Aoyama:2012, Aoyama:2019}.  In practice the full five-loop series is used numerically when producing Eqs.~\eqref{eq:deltaaeRb}--\eqref{eq:deltaaeCs}.  The sign flip between the two $\Delta a_e$ values arises because the Rb determination yields a slightly \emph{larger} $\alpha^{-1}$ (smaller $\alpha$) than the Cs value, shifting the QED prediction and thus the residual.
%\paragraph{Implications for New Physics.}
%Any new light state (e.g.\ a kinetically mixed dark photon) contributing a positive shift to $a_e$ would help explain a positive $\Delta a_e$ but exacerbate a negative one.  Consequently, model fits often consider both determinations separately, constructing allowed or excluded regions in $(m_V, \varepsilon)$ space using
%\begin{equation}
%  a_e^{\rm NP}(m_V,\varepsilon) = \frac{\alpha\, \varepsilon^2}{2\pi} F_V\!\left(\frac{m_V}{m_e}\right),
%\end{equation}
%with the loop function $F_V$ given in Sec.~\ref{sec:loopfunction} (cf.\ \cite{Pospelov:2008zw, Bjorken:2009mm}).  Bounds are then derived by requiring $|a_e^{\rm NP}|\leq |\Delta a_e^{(i)}| + n\sigma$ for a chosen confidence level, using separately Eq.~\eqref{eq:deltaaeRb} and Eq.~\eqref{eq:deltaaeCs}.

The two precision $\alpha$ measurements lead to electron anomaly residuals of opposite sign, Eqs.~\eqref{eq:deltaaeRb}--\eqref{eq:deltaaeCs}.  Any global fit to new-physics explanations of magnetic moment data must therefore treat the Cs 2018 and Rb 2020 inputs as distinct scenarios.
%\end{flushleft}

%% file: heavy_photon_mixing.tex
\section{Heavy Photon mixing with a lepton at Vacuum level}

\begin{figure}[ht]
\centering
\begin{tikzpicture}[>=stealth, thick,
    photon/.style={decorate, decoration={snake, amplitude=1pt, segment length=4pt}, thick},
    aprime/.style={dashed, thick}
]
  %---- Coordinates ----
  \coordinate (top) at (0,2);       % photon vertex
  \coordinate (left) at (-1.5,-1);  % muon in
  \coordinate (right) at (1.5,-1);  % muon out
  \coordinate (midleft) at ($(left)!0.55!(top)$);   % attach A'
  \coordinate (midright) at ($(right)!0.55!(top)$); % attach A'

  %---- Muon lines ----
  \draw[->] (left) -- (top) node[midway,left] {$\ell^-$};
  \draw[->] (top) -- (right) node[midway,right] {$\ell^-$};

  %---- External photon ----
  \draw[photon] (top) -- ++(0,1.5) node[midway,right] {$\gamma$};

  %---- A' leading-order contribution ----
  \draw[aprime] (midleft) -- (midright) node[midway,below] {A$^\prime$};

\end{tikzpicture}
\caption{Leading-order heavy photon contribution to the gyromagnetic ratio of a lepton.}
\label{fig:leading_aprime_muon}
\end{figure}
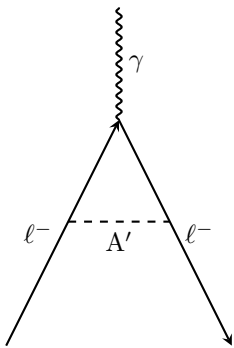

\subsection{Calculating Leptonic Kinetic Mixing Parameters}

In the minimal kinetic mixing (``dark/heavy photon'') scenario with Lagrangian
\[
\mathcal{L} \supset -\frac{1}{4}F'_{\mu\nu}F'^{\mu\nu} - \frac{\varepsilon}{2}F'_{\mu\nu} B^{\mu\nu} + \frac{1}{2} m_{A'}^{2} A'_\mu A'^{\mu}
\]
the induced interaction after electroweak symmetry breaking is flavor universal,
\[
\mathcal{L}_{\rm int} \supset \varepsilon e \sum_{f} Q_f \bar f \gamma^\mu f\, A'_\mu,
\]
thus the same kinetic mixing $\varepsilon$ governs both (i) the one–loop correction to $a_\mu$ and $a_e$ and (ii) production / decay into $e^+e^-$ relevant for the nuclear transition ATOMKI anomaly.

The one–loop contribution of a light vector of mass $m_{A'}$ to a charged lepton anomalous moment is calculated using the form factor $F_V$ where 
\[
\Delta a_\ell^{A'} = \frac{\alpha \varepsilon^2}{2\pi}\, F_V\!\left(\frac{m_{A'}}{m_\ell}\right),
\qquad
F_V(r)=\int_{0}^{1} dx\, \frac{2x(1-x)^2}{(1-x)^2 + r^{2}x},
\]
which yields the limiting forms $F_V(r\!\ll\!1)\simeq 1$ and $F_V(r\!\gg\!1)\simeq \frac{2}{3}\frac{m_\ell^{2}}{m_{A'}^{2}}$, allowing a direct mapping between a measured (or hypothesized) $\Delta a_\ell$ and the required $\varepsilon$ \cite{Pospelov:2008zw}.

Thus, the kinetic mixing parameter is

\begin{equation}
\varepsilon_\ell(m_{A^\prime}) = \sqrt{ \frac{ \Delta a_\ell \left( 2\pi / \alpha \right) }{ F_V\left(m_{A^\prime}{} / m_\ell\right) } }
\end{equation}

%where the form factor $f(r = \frac{m_{A^{\prime}}}{m_\mu})$ is given by:
%\begin{equation}
%f(r) = \int_0^1 \frac{2z(1 - z)^2}{(1 - z)^2 + r^2 z} dz
%\end{equation}

Here, $m_\mu$ is the mass of the respective lepton, and $\alpha$ is the fine structure constant ($\alpha \approx 1/137$). The $1\sigma$ uncertainty on the kinetic mixing parameter $\epsilon_\ell$ is propagated as:
\begin{equation}
\sigma_{\epsilon_\ell} = \frac{1}{2} \frac{\epsilon_\ell}{\Delta a_\ell} \sigma_{\Delta a_\ell}
\end{equation}
where $\Delta a_\ell$ and $\sigma_{\Delta a_\ell}$ are the central value and uncertainty of the anomalous magnetic moment discrepancy, respectively.

At next-to-leading order the dark photon contribution acquires standard QED two–loop vertex and self-energy corrections. These dress the one–loop amplitude without introducing additional powers of $\epsilon$. The corrected expression can be written
\[
\Delta a_\ell^{A'} = \frac{\alpha \epsilon^2}{2\pi} F_V(r)\Bigl[1 + \frac{\alpha}{\pi}\delta_V(r)\Bigr],
\]
where $F_V$ is given above and $\delta_V(r)=\mathcal{O}(1)$ encodes the finite two–loop terms (no large logarithms for $m_{A'}\sim \mathcal{O}(m_\ell)$).

\[
\frac{\Delta a_\ell^{A',\text{NLO}}-\Delta a_\ell^{A',\text{LO}}}{\Delta a_\ell^{A',\text{LO}}} \sim \mathcal{O}\!\left(\frac{\alpha}{\pi}\right) \approx 2.3\times 10^{-3}
\]

Numerically, this modifies $\epsilon$ extracted from $\Delta a_\ell$ at the $10^{-3}$ relative level, negligible compared to the leading order uncertainties.  Thus, leading order expressions suffice for the parameter-space contours presented here.

%\subsection{Uncertainty Propagation}
%The $1\sigma$ uncertainty on the kinetic mixing parameter $\epsilon_\mu$ is propagated as:
%\begin{equation}
%\sigma_{\epsilon_\mu} = \frac{1}{2} \frac{\epsilon_\mu}{\Delta a_\mu} \sigma_{\Delta a_\mu}
%\end{equation}
%where $\Delta a_\mu$ and $\sigma_{\Delta a_\mu}$ are the central value and uncertainty of the anomalous magnetic moment discrepancy, respectively.

%% file: results.tex
\section{Results}

\subsection{Determining $\varepsilon_\mu$ using $\Delta a_\mu$}

%\begin{table}[h!]
%\centering
%\begin{tabular}{@{}cccc@{}}
%\toprule
%\textbf{Mass [MeV]} & \textbf{$\varepsilon_{\text{disp}}$} & \textbf{$\varepsilon_{\text{BMW}}$} & \textbf{$\varepsilon_{2025}$} \\
%\midrule
%5   & $1.57\times10^{-3}$ & $1.03\times10^{-3}$ & 111\\
%10  & $1.66\times10^{-3}$ & $1.09\times10^{-3}$ & 111\\
%15  & $1.75\times10^{-3}$ & $1.14\times10^{-3}$ & 111\\
%17  & $1.78\times10^{-3}$ & $1.17\times10^{-3}$ & 111\\
%20  & $1.84\times10^{-3}$ & $1.20\times10^{-3}$ &111\\
%25  & $1.92\times10^{-3}$ & $1.25\times10^{-3}$ &111\\
%30  & $2.00\times10^{-3}$ & $1.31\times10^{-3}$ &111\\
%35  & $2.09\times10^{-3}$ & $1.36\times10^{-3}$ &111\\
%40  & $2.17\times10^{-3}$ & $1.42\times10^{-3}$ &111\\
%45  & $2.25\times10^{-3}$ & $1.47\times10^{-3}$ &111\\
%50   & $2.25\times10^{-3}$ & $1.47\times10^{-3}$  &111\\
%100  & $2.25\times10^{-3}$ & $1.47\times10^{-3}$ &111\\
%200  & $2.25\times10^{-3}$ & $1.47\times10^{-3}$ &111\\
%300  & $2.25\times10^{-3}$ & $1.47\times10^{-3}$ &111\\
%400  & $2.25\times10^{-3}$ & $1.47\times10^{-3}$ &111\\
%500  & $2.25\times10^{-3}$ & $1.47\times10^{-3}$ &111\\
%\bottomrule
%\end{tabular}
%\caption{Kinetic mixing $\varepsilon$ values for masses $m_X \in [5, 500]$ MeV under both dispersive and lattice-QCD corrected $(g-2)_\mu$ scenarios. [TODO --- numbers for all $\varepsilon_{2025}$, double check other numbers, fill in correct numbers for 50-500 MeV]}
%\end{table}

%\subsection*{$\varepsilon_\mu$}
Following the prescription of \textbf{Section 3.1} we calculate the kinetic mixing parameter, $\varepsilon_{\mu}$, for heavy photon masses, $m_X$ within the range $[5, 500]$ MeV for three theoretical scenarios. The calculated values for a subset within this range are listed in Table 1 and explicitly plotted in \ref{fig:g2_band_comparison}.

As an example, and to compare with the ATOMKI measurement, suppose $m_{A^{\prime}} = 17\,\mathrm{MeV}$. We have $r \approx 0.161$ and $f(r) \approx 0.68$. Taking the latest values:
\begin{align*}
\Delta a_\mu^{\mathrm{disp}} &= 262 (45) \times 10^{-11} \quad \text{(WP 2020)} \\
\Delta a_\mu^{\mathrm{BMW}}  &= 118 (69)\times 10^{-11} \quad \text{(BMW 2021)}
\end{align*}
we obtain:
\[
\varepsilon_\mu^{\mathrm{disp}} = \sqrt{ \frac{ 262 \times 10^{-11} \cdot (2\pi/ \alpha ) }{ 0.68 } }
\approx 1.82 \times 10^{-3}
\]
and
\[
\varepsilon_\mu^{\mathrm{BMW}} = \sqrt{ \frac{ 118 \times 10^{-11} \cdot (2\pi/ \alpha) }{ 0.68 } }
\approx 1.22 \times 10^{-3}
\]
Therefore, the original muon-$(g-2)$–favored kinetic mixing,
\[
\varepsilon_\mu^{\mathrm{disp}} \approx 1.82 \times 10^{-3},
\]
shrinks to
\[
\varepsilon_\mu^{\mathrm{BMW}} \approx 1.22 \times 10^{-3}
\]
once BMW lattice corrections are included, moving the central value of the allowable coupling strength within the ATOMKI X17 preferred range of
\[
2\times10^{-4} \leq |\varepsilon| \leq 1.4\times10^{-3}
\]
for a $16.7\,\mathrm{MeV}$ boson~\cite{PhysRevLett.117.071803}.  The (g-2) theory initiative 2025 white paper then narrows the central value to the allowable coupling further, where  
\[
\Delta a_\mu^{\rm WP25} = 39(64) \times 10^{-11}
\]
corresponding to 
\[
\varepsilon_\mu^{\mathrm{WP25}} = 
7.034 \times10^{-4}
\]

\begin{figure}
    \centering
    \includegraphics[width=\linewidth]{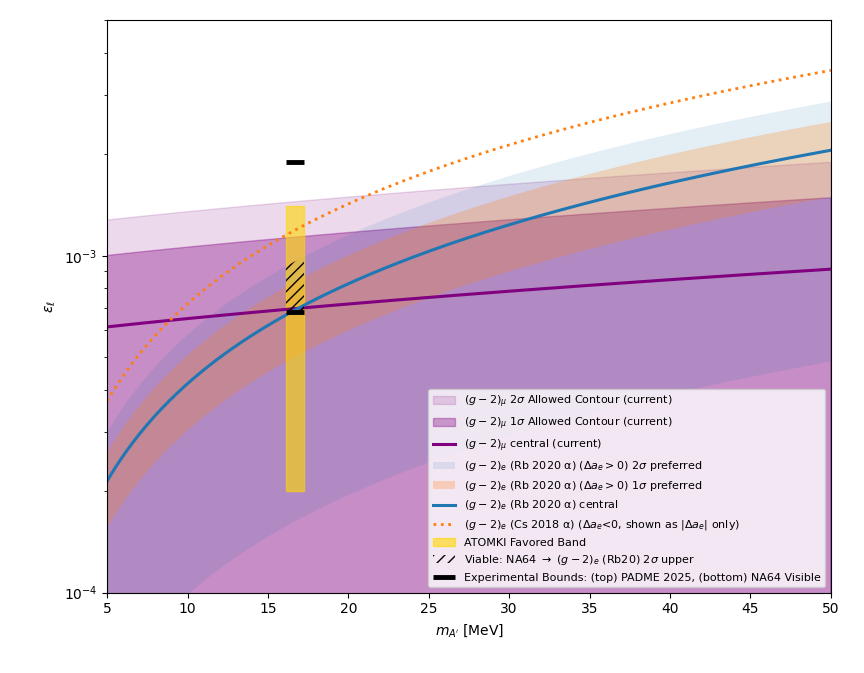}
    \caption{Overlay of ATOMKI preferred $\varepsilon_e$ (yellow box) and allowable $1\sigma$, $2\sigma$ contours on $\varepsilon_e$, $\varepsilon_\mu$ from fine structure precision measurements on the $^{87}$Rb, $^{133}$Cs and the FNAL final average $a_\mu$ respectively. The black lines indicate the current limits set by (top) PADME, and (bottom) NA64. In between the NA64 limit and the top of the 2$\sigma$ favored $\varepsilon^{Rb20}_e$ band is the region still viable for the protophobic and neutrino-coupling suppressed interpretation of X17.}
    \label{fig:curved_bands}
\end{figure}

 These three theoretical predictions for allowed coupling on $\varepsilon$ with corresponding $1 \sigma$ and $2 \sigma$ contours are overlayed with the ATOMKI X17 parameter band in Figure \ref{fig:curved_bands}. Table 1 provides the computations for a subset of hypothetical vector boson masses within the range [5, 500] MeV. 

\begin{table}[ht]
\centering
\label{tab:eps_scenarios}
\begin{tabular}{rcccccc}
\toprule
 {Mass [MeV]} & {\(\varepsilon_{\text{WP20}}\)} & {\(\sigma(\varepsilon_{\text{WP20}})\)} &
 {\(\varepsilon_{\text{BMW}}\)} & {\(\sigma(\varepsilon_{\text{BMW}})\)} &
 {\(\varepsilon_{\text{WP25}}\)} & {\(\sigma(\varepsilon_{\text{WP25}})\)} \\
 % & & & & & & \\
\midrule
  5   & 1.604e-03 & 1.377e-04 & 1.076e-03 & 3.146e-04 & 6.187e-04 & 5.077e-04 \\
 10   & 1.698e-03 & 1.458e-04 & 1.139e-03 & 3.331e-04 & 6.549e-04 & 5.374e-04 \\
 15   & 1.788e-03 & 1.535e-04 & 1.200e-03 & 3.508e-04 & 6.897e-04 & 5.659e-04 \\
 17   & 1.823e-03 & 1.566e-04 & 1.223e-03 & 3.577e-04 & 7.034e-04 & 5.771e-04 \\
 20   & 1.875e-03 & 1.611e-04 & 1.259e-03 & 3.680e-04 & 7.236e-04 & 5.937e-04 \\
 25   & 1.962e-03 & 1.685e-04 & 1.316e-03 & 3.849e-04 & 7.568e-04 & 6.210e-04 \\
 30   & 2.047e-03 & 1.758e-04 & 1.373e-03 & 4.016e-04 & 7.896e-04 & 6.479e-04 \\
 35   & 2.131e-03 & 1.830e-04 & 1.430e-03 & 4.181e-04 & 8.220e-04 & 6.745e-04 \\
 40   & 2.214e-03 & 1.901e-04 & 1.486e-03 & 4.344e-04 & 8.542e-04 & 7.009e-04 \\
 45   & 2.297e-03 & 1.972e-04 & 1.541e-03 & 4.507e-04 & 8.862e-04 & 7.271e-04 \\
 50   & 2.379e-03 & 2.043e-04 & 1.597e-03 & 4.668e-04 & 9.180e-04 & 7.532e-04 \\
100   & 3.192e-03 & 2.741e-04 & 2.142e-03 & 6.263e-04 & 1.231e-03 & 1.010e-03 \\
200   & 4.812e-03 & 4.132e-04 & 3.229e-03 & 9.441e-04 & 1.856e-03 & 1.523e-03 \\
300   & 6.448e-03 & 5.538e-04 & 4.327e-03 & 1.265e-03 & 2.488e-03 & 2.041e-03 \\
400   & 8.101e-03 & 6.957e-04 & 5.437e-03 & 1.590e-03 & 3.126e-03 & 2.565e-03 \\
500   & 9.768e-03 & 8.389e-04 & 6.555e-03 & 1.917e-03 & 3.769e-03 & 3.092e-03 \\
\bottomrule
\end{tabular}
\caption{Calculated kinetic mixing parameter $\varepsilon_\mu$ (and $1\sigma$ uncertainties) for three scenarios [dispersive only (``WP20''), BMW lattice 2021 calculations (``BMW''), and 2025 white paper (``WP25'')]using the final FNAL average $a_\mu^\mathrm{exp}$, for a subset of hypothetical vector boson masses within the range $[5, 500]$ MeV.} 
\end{table}

\subsection{Determining $\varepsilon_e$ using $\Delta a_e$}

Although we don't explicitly make a table for a range of selections that could be allowable for $\Delta a_e$, the central values for $\varepsilon_e$ when $m_{A^{\prime}}=17$ MeV are.
\begin{equation}
    \varepsilon^{\text{CS18}}_e = 1.19 (15) \times 10^{-3} 
\end{equation}
\begin{equation}
    \varepsilon^{\text{RB20}}_e = 0.69 (15)\times 10^{-3} 
\end{equation}

\begin{figure}[h!]
    \centering
    \begin{subfigure}{0.55\linewidth}
        \centering
        \includegraphics[width=\linewidth]{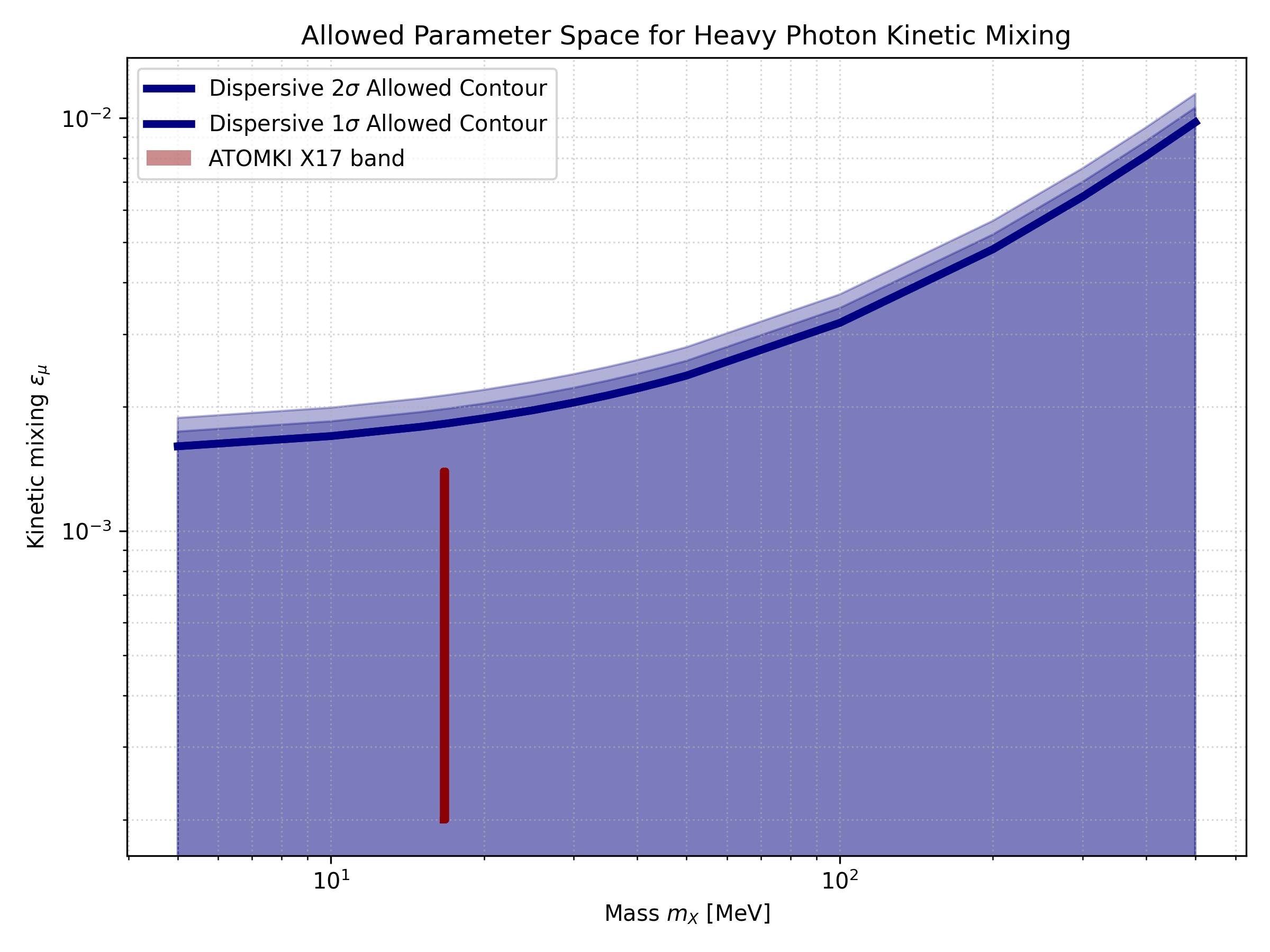}
        \caption{}
        \label{fig:curved_bands1}
    \end{subfigure}
    
    \begin{subfigure}{0.55\linewidth}
        \centering
        \includegraphics[width=\linewidth]{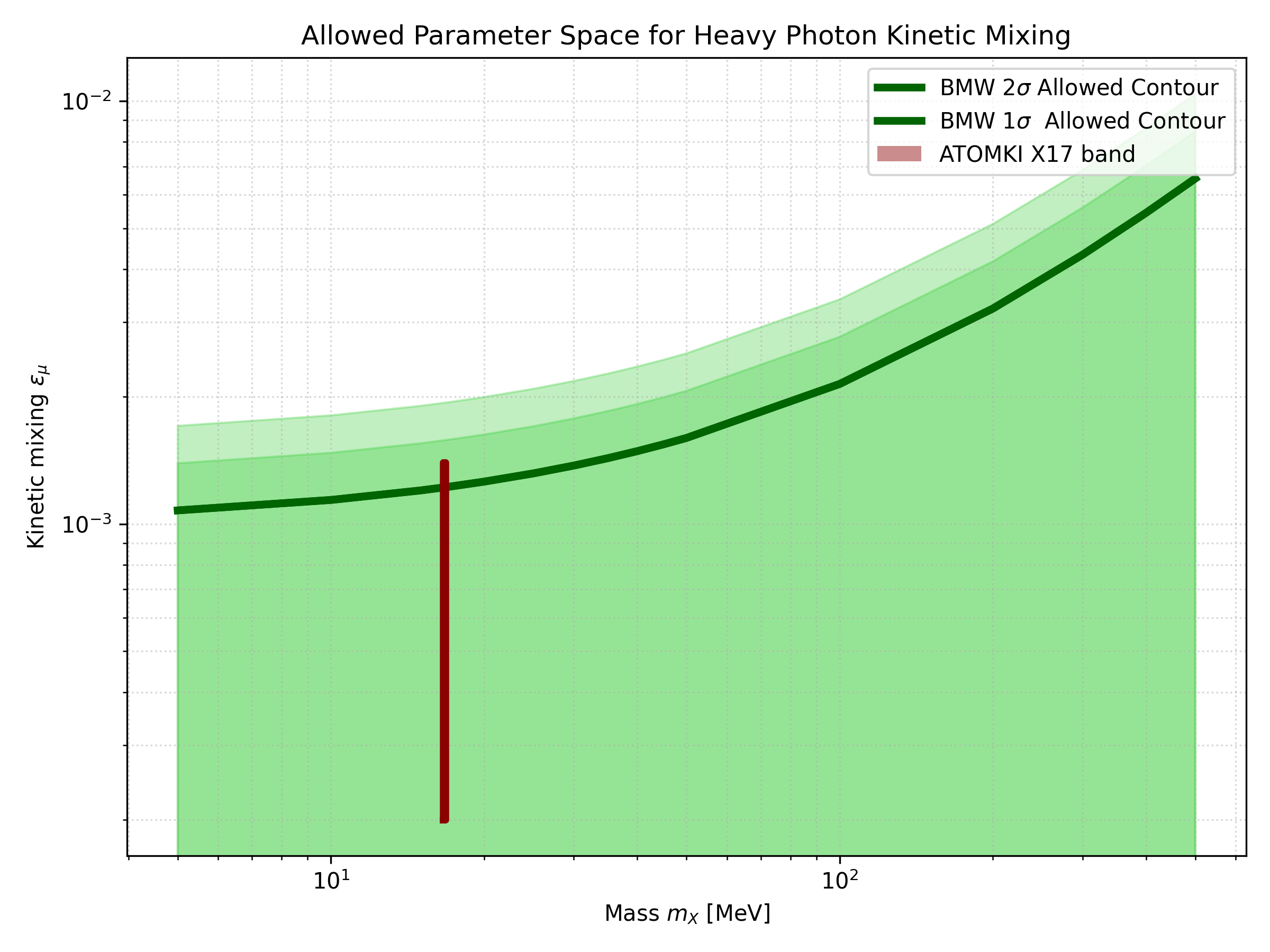}
        \caption{}
        \label{fig:curved_bands2}
    \end{subfigure}
    
    \begin{subfigure}{0.55\linewidth}
        \centering
        \includegraphics[width=\linewidth]{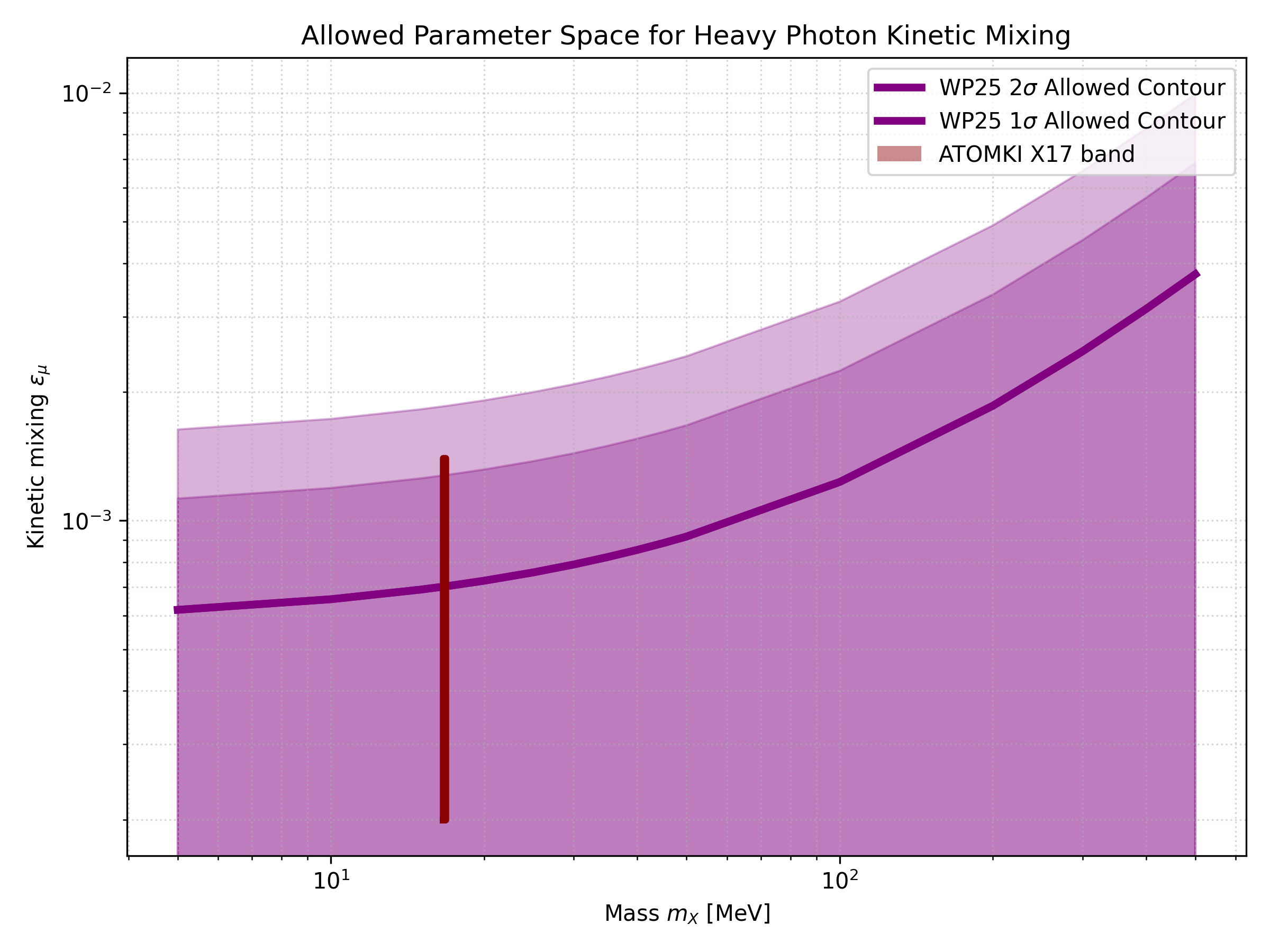}
        \caption{}
        \label{fig:curved_bands3}
    \end{subfigure}

    \caption{Comparison of $(g-2)_\mu$ allowed parameter space using: WP20 (a), BMW21 (b), and WP25 (c) theory predictions.}
    \label{fig:g2_band_comparison}
\end{figure}

Given the vector intepretation necessitating a positive value of $\Delta a_e$, we only plot the mean $\varepsilon^{CS18}_e$ (using $|\Delta a_e|$) alongside the 1$\sigma$ and 2$\sigma$ favored regions about $\varepsilon^{RB20}_e$ as calculated for masses $m_{A^{\prime}}\in [5, \;50]$ MeV. This is plotted alongside the 1$\sigma$ and 2$\sigma$ allowed $\varepsilon_\mu$ and the ATOMKI/NA64 observed limits in \ref{fig:g2_band_comparison}.

%\begin{table}[h!]
%\centering
%\begin{tabular}{lcccc}
%\toprule
%Measurement & $\Delta a_\mu$ & Central $\epsilon$ & Upper Uncertainty & Lower Uncertainty \\
%\midrule
%Dispersive (2023) &1111 (10) & $2.154\times10^{-3}$ & $+2.398\times10^{-4}$ & $-2.701\times10^{-4}$ \\
%BMW lattice (2023) & 11111 &$1.406\times10^{-3}$ & $+2.476\times10^{-4}$ & $-3.018\times10^{-4}$ \\
%FNAL 2023 & 11111 & $2.154\times10^{-3}$ & $+2.398\times10^{-4}$ & $-2.701\times10^{-4}$ \\
%FNAL 2025 Dispersive & 11111 & $2.154\times10^{-3}$ & $+2.398\times10^{-4}$ & $-2.701\times10^{-4}$ \\
%FNAL 2025 BMW & 1111112 & $1.406\times10^{-3}$ & $+2.476\times10^{-4}$ & $-3.018\times10^{-4}$ \\
%\bottomrule
%\end{tabular}
%\caption{$\Delta a_\mu$ and corresponding kinetic mixing parameters ($\epsilon$) with uncertainties.}
%\end{table}

%\begin{figure}
%    \centering
%    \includegraphics[width=0.9\linewidth]{g2_parameter_space_WP20.png}
%    \caption{}
%    \label{fig:curved_bands1}
%\end{figure}

%\begin{figure}
%    \centering
%    \includegraphics[width=0.9\linewidth]{g2_parameter_space_BMW21.png}
%    \caption{}
%    \label{fig:curved_bands2}
%\end{figure}

%\begin{figure}
%    \centering
%    \includegraphics[width=0.9\linewidth]{g2_parameter_space_WP25.png}
%    \caption{}
%    \label{fig:curved_bands3}
%\end{figure}

%\begin{figure}[ht!]
%    \centering
%    \includegraphics[width=0.9\textwidth]%{updated_g2_final_final.png}
%    \caption{Kinetic mixing parameter comparison clearly showing the ATOMKI X17 region ($16.6$--$16.8\,\mathrm{MeV}$, $2\times10^{-4}$--$1.4\times10^{-3}$, shaded red) alongside multiple confirmed theoretical and experimental $(g-2)$ results.}
%    \label{fig:main}
%\end{figure}

%Need two sigma contor and central line through the $\varepsilon^2$ parameter space

%% file: best_conclusion.tex
\section{Conclusion}
This work re–evaluates whether a kinetically mixed dark photon, $A'$, can reconcile the muon and electron magnetic–moment anomalies with the nuclear $e^{+}e^{-}$ excess reported by ATOMKI at $17,\text{MeV}$.  The final 2025 Fermilab $(g-2)_\mu$ result, together with the BMW21 lattice and WP25 theory updates, plays a central role in reassessing the favored kinetic–mixing parameter $\epsilon$.

\paragraph{Impact of the 2025 $(g-2)_\mu$ determination.}  Using the same experimental input but three successive theory treatments, the preferred muonic coupling has moved steadily downward:
\begin{equation}
\varepsilon_\mu^{\text{WP20}} = 1.82\times10^{-3},\qquad
\varepsilon_\mu^{\text{BMW21}} = 1.22\times10^{-3},\qquad
\varepsilon_\mu^{\text{WP25}} = 7.03\times10^{-4}.
\end{equation}
The reduction reflects improved control of hadronic vacuum–polarisation contributions; each refinement narrows the viable parameter space by roughly a factor of two.

For the electron, the one-electron quantum cyclotron combined with the Cs and Rb recoil determinations of the fine–structure constant leads to a $(g-2)_e$ residual that prefers a slightly smaller mixing with the following means:
\begin{equation}
\varepsilon^{\text{CS18}}_e = 1.19  \times 10^{-3},\qquad
\varepsilon^{\text{RB20}}_e = 6.91 \times 10^{-4}
\end{equation}
We recognize that the CS18 value is not to be considered in preferred coupling given the negative value of $\Delta a_e$ (as vector bosons necessarily have a positive contribution to the ratio) and thus conclude with the RB20 value. At the X17 mass, the Rb20 electron preferred band and the muon exclusion band still overlaps at the $2\sigma$ level, so a universal (or mildly non‑universal) $A^{\prime}$ interpretation remains possible.

Overlaying the leptonic constraints with the ATOMKI signal region yields the viable search range of 
\begin{equation}
6.8\times10^{-4}  \lesssim  \epsilon  \lesssim  9.6\times10^{-4},
\end{equation}
where the bottom of the range is given by NA64 and the max value of the range corresponds to the 2$\sigma$ upper bound consistent with the favored electron coupling region from the 2020 Rb measurement. This region remains consistent with a protophobic gauge boson that satisfies the NA48/2 $\pi^{0}\to\gamma A'$ bound.  The window is narrow but not yet excluded.

%\paragraph{Impact of the 2025 $(g-2)_\mu$ determination.}
%Several forthcoming data sets will probe this region more stringently.  A dedicated X17 experiment at Jefferson Lab, scheduled for late 2025, will directly test the $17,\text{MeV}$ hypothesis.  Continued running of the Fermilab storage‑ring experiment will further reduce the uncertainty in $(g-2)_\mu$, while the J‑PARC ultra‑cold muon $(g-2)/$EDM experiment will provide a cross‑check.  On the electron side, next‑generation recoil measurements aim to improve the precision on $\alpha$ and hence on $(g-2)_e$. 

PIONEER, Mu3e, and X17@JLAB are positioned to probe the remaining viable parameter space. The results of these programs will determine if anything  in this parameter space corresponds to physics beyond the Standard Model or must be ruled out.  In either case, the forthcoming results will decisively clarify the role of light vector bosons in the sub‑GeV sector and test the validity of the ATOMKI claims.